\documentclass{PoS}
\usepackage{amsmath}
\usepackage{cite}

\usepackage{tikz}

\newcommand\encircle[1]{  \tikz[baseline=(X.base)] \node (X) [draw, shape=circle, inner sep=0] {\strut #1};}

\newcommand{\sm}{{Standard Model }}

\def\lfv{lepton flavour violation }
\def\lnv{lepton number violation }

\def\O{$\cal O$ }
\def\vev#1{\left\langle #1\right\rangle}

\def\TrTrTrOne{$\mathrm{SU(3)_C \otimes SU(3)_L \otimes SU(3)_R \otimes U(1)_X }$ }
\def\TrTrOne{$\mathrm{SU(3)_C \otimes SU(3)_L \otimes U(1)_X }$ }

\def\SM{$\mathrm{ SU(3)_C \otimes SU(2)_L \otimes U(1)_Y }$ }
\def\LR{$\mathrm{SU(3) \otimes SU(2)_L \otimes SU(2)_R \otimes U(1)_{B-L} }$ }

\def \nbb {$\beta\beta_{0\nu}$ }

\newcommand{\AddrAHEP}{AHEP Group, Institut de F\'{i}sica Corpuscular --
  C.S.I.C./Universitat de Val\`{e}ncia \\ Parc Cientific de Paterna.
  C/Catedratico Jos\'e Beltr\'an, 2 E-46980 Paterna (Val\`{e}ncia) - SPAIN}

\title{ Neutrino physics  from A to Z : Two lectures at Corfu} 

\ShortTitle{ Neutrino physics from A to Z : two lectures at Corfu}

\author{J.W.F. Valle\\
{\AddrAHEP} 
E-mail: \email{valle@ific.uv.es, URL:  http://astroparticles.es/}}

\abstract{Assuming basic familiarity with neutrino physics, I give a
  telegraphic and panoramic view of the current status and the main
  open challenges in the field. I also illustrate how the mechanism
  responsible for neutrino mass generation may shed light upon some of
  the current puzzles in particle physics as well as
  cosmology\footnote{This writeup corresponds to a slightly
    modified/updated version of the material actually presented at the
    School.}.  }

\FullConference{Proceedings of the Corfu Summer Institute 2016\\  
"Summer School and Workshop on the Standard Model and Beyond"\\
  August 31 - September 12, 2016\\
  Corfu, Greece}

\begin{document}

\noindent
\textbf{\encircle{A}  Introduction. } 
The last few decades have seen a tremendous progress in particle
physics and cosmology, the physics of the early universe. The
discoveries of the Higgs boson at the Large Hadron Collider at
CERN~\cite{Aad:2012tfa,Chatrchyan:2012ufa} and of neutrino
oscillations~\cite{Kajita:2016cak,McDonald:2016ixn} as a result of
solar and atmospheric studies constitute major milestones in
astroparticle physics which led to the 2012 and 2015 physics Nobel
prizes.
It is not an overstatement to say that the discovery of the Higgs
boson has brought our field to euphoria. Some considered it as the
last brick in the construction of the standard model. It is not, since
in the \sm neutrinos have no mass, needed to account for the
oscillation data~\cite{Forero:2014bxa}. More bricks are needed to
fabricate neutrino mass.
Indeed, as a key building block of the Standard Model, the properties
of neutrinos may point us the way beyond the standard
model~\cite{Valle:2015pba}. However small the magnitude of their
masses results to be, the electroweak breaking mechanism can be
significantly affected by the presence of massive neutrinos, with
potentially profound implications, e.g. for the consistency of the
electroweak vacuum.
Likewise neutrinos constitute an ideal cosmic messenger capable of
exploring the earliest moments after the Big Bang.
For example, establishing the existence of leptonic CP violation is an
important goal in the agenda of upcoming oscillation experiments such
as that of the DUNE proposal.
Such a discovery would pave the way to elucidate one of the great
cosmological puzzles i.e. the understanding the prevalence of matter
over anti-matter in our universe.

On the other hand, the \sm does not include gravity.  The first
observation of gravitational waves by the LIGO Scientific
Collaboration and Virgo Collaboration
teams~\cite{PhysRevLett.116.061102}, have brought the ultimate need
for the inclusion of gravity in our world picture more into the
forefront than ever. Reconciling gravity and quantum mechanics is a
formidable challenge that lies with us for over a century.  Now the
time seems to have come ! 
There are, in addition, a variety of other, theoretical motivations
for having beyond the \sm physics, such as understanding anomaly
cancellation, unification of the forces, the consistency of the
spontaneous symmetry breaking mechanism, including naturalness,
stability and perturbativity. 
Unfortunately, other than the discovery of neutrino mass and
some cosmological hints, the search for unambiguous signs of new
phenomena in particle physics has so far been fruitless.
In these lectures I will assume that you know the basics and focus on
illustrating how the theory responsible for generating neutrino mass
may also touch several of the above points. 

\noindent
\textbf{\encircle{B}  Neutrino probes. }
Neutrinos are tiny weakly interacting particles travelling close to
the speed of light. They constitute one of the most ubiquitous
particles in nature.  Thanks to their weak interaction, neutrinos are
excellent astrophysical messengers, probing the deep interior of the
Sun or of a supernova.  Likewise they probe the earliest instants of
the universe, just after the Big Bang.
Natural and artificial neutrino sources span about 20 orders of
magnitude in energy, all the way from the abundant neutrinos produced
in the Big-Bang to the ultra high energy cosmic ray neutrinos. The
former are abundant, though currently undetectable because of their
very low energy.  The latter have much higher interaction rates,
though their detection also constitutes a challenge thanks to the
small fluxes.
In between these extremes we have geoneutrinos, supernova
neutrinos, as well as solar and atmospheric neutrinos.\\[-.1cm]

\noindent
\textbf{\encircle{C}  Solar and atmospheric neutrinos. }
Neutrinos are copiously produced in nuclear fusion reactions that
power the Sun, while atmospheric neutrinos hit the Earth from all
directions in the sky, produced by cosmic ray interaction in the upper
atmosphere.
These neutrinos are detected in gigantic underground detectors.
In both cases there is a significant discrepancy between produced and
detected neutrinos.
The study of solar and atmospheric neutrinos wrote an important
chapter in particle physics, leading to the discovery of neutrino
oscillations~\cite{Kajita:2016cak,McDonald:2016ixn} and the physics
Nobel prize in 2015. The discovery of neutrino oscillation has been
beautifully confirmed by earthbound experiments based at reactors and
accelerators.

\begin{figure}[h]
\centering
\includegraphics[width=0.475\textwidth]{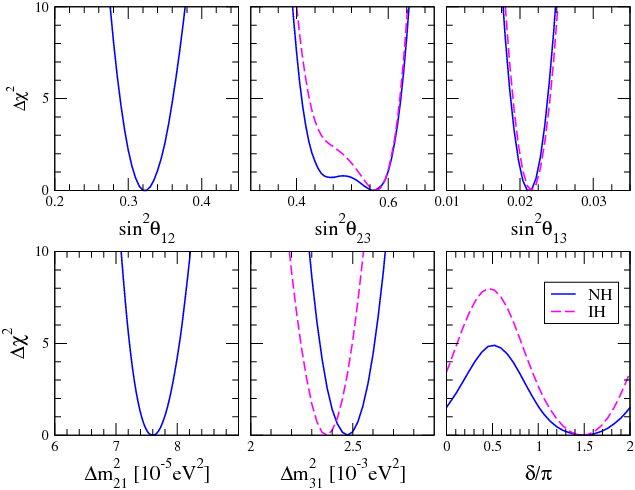}
\includegraphics[width=0.4\textwidth,height=5.5cm]{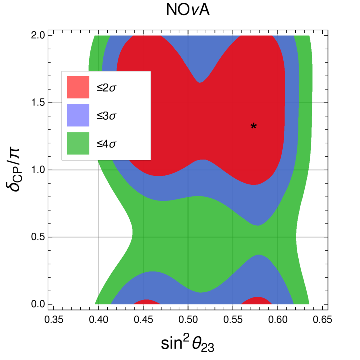}
\caption{Left: Current status of oscillation parameters. Note the
  octant ambiguity in the atmospheric angle determination for NH and
  the very poor CP violating phase determination, according to the
  global fit \protect\cite{Forero:2014bxa}. Right: expected octant and
  CP sensitivity at the long baseline oscillation experiment NOvA,
  from~\protect\cite{Chatterjee:2017xkb}.}
\label{ref:global}
\end{figure}

\noindent
\textbf{\encircle{D}  Reactors and accelerator neutrinos. }
All of these data are beautifully described by assuming that neutrinos
undergo oscillations, a quantum mechanical phenomenon, as they
propagate.
The phenomenon is affected by the presence of matter, as it happens
both in the interior of the Sun as well as in the Earth.
Except for the atmospheric angle and CP violating phase, current
neutrino oscillation experiments provide a good determination of the
oscillation parameters, as seen in Fig.~\ref{ref:global}.
Apart from the smallness of neutrino squared mass splittings, one thing
that strikes the eye is the large values of the lepton mixing angles
with respect to those that characterize the Kobayashi-Maskawa matrix.
Resolving the atmospheric octant will require an improved measurement
of the reactor angle $\theta_{13}$~\cite{Chatterjee:2017xkb}.
The precision in the measurement of $\theta_{23}$ and $\delta_{CP}$
will be improved at the long baseline oscillation experiment NOvA and
at the upcoming DUNE The bands correspond to the 2, 3, and 4$\sigma$
C.L uncertainties.
Note that oscillations do not probe the absolute neutrino mass nor are
they currently sensitive to the ordering of the neutrino states.\\[-.1cm]

\begin{figure}[h]
\centering
\includegraphics[width=0.47\textwidth,height=5cm]{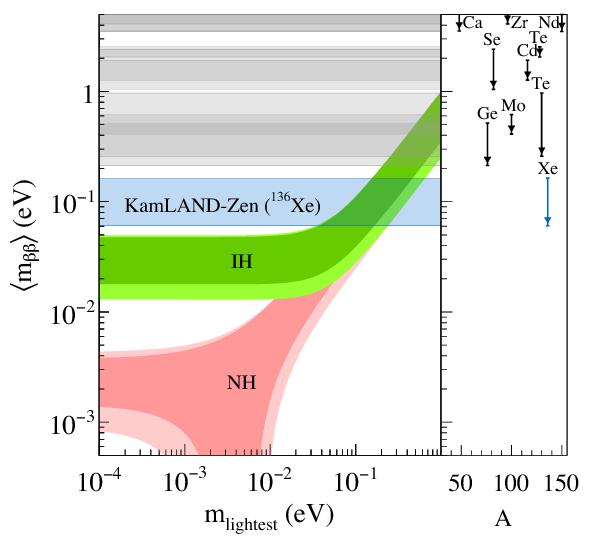}
\caption{Status and prospects for \nbb searches,
  from~\cite{KamLAND-Zen:2016pfg}, see text.  }
\label{ref:bbglobal}
\end{figure}

\noindent
\textbf{\encircle{E}  The absolute neutrino mass. } 
Currently there are three realistic ways to probe the absolute
neutrino mass: (i) measuring the shape of the end point of the
spectrum in tritium beta decays, (ii) the search for neutrinoless
double beta decay \nbb as well as (iii) measurements of temperature
anisotropies in the cosmic microwave background.
Now we discuss (ii). Two-neutrino double-beta decay is the
second-order weak interaction process by which two neutrons in a
nucleus are converted to protons, plus two electrons plus two electron
anti-neutrinos, $(A, Z) \to (A, Z + 2) + 2 e + 2\nu$. This very rare
process has been detected in a few nuclei. It conserves lepton number.
On the other hand \nbb is a neutrinoless variety expected to occur if
neutrinos are Majorana. Its amplitude is proportional to an effective
mass parameter $\vev{m_{\beta\beta}}$ given in Fig.~\ref{ref:bbglobal}
as a function of the lightest neutrino mass. The dark shaded regions
are generic predictions based on best-fit neutrino oscillation
parameters for normal hierarchy (NH) and inverted hierarchy (IH). The
light shaded regions are the corresponding $ 3\sigma$ ranges. The
lowest horizontal band indicates the 90\% C.L. upper limit on
$\vev{m_{\beta\beta}}$ from KamLAND-Zen, using $^{136}$ Xe.  The upper
bands give the sensitivities for other nuclei taking into account
nuclear matrix element calculations~\cite{KamLAND-Zen:2016pfg}. The
side-panel shows the corresponding limits for each nucleus as a
function of the mass number.
One sees a big experimental race to search \nbb and possibly discover
it! However, in the case of NH there can be destructive interference
leading to a cancellation in the expected \nbb rate. 
From this point of view it is very interesting to consider theories
where the flavor structure is predicted in such a way that the
cancellation is prevented. In such models there is a lower bound on
the \nbb decay rate. Examples are given
in~\cite{Dorame:2011eb,Dorame:2012zv,King:2013hj,Bonilla:2014xla}.
Note also that \nbb can be induced by a short range mechanism,
mediated by heavy states, through a variety of
mechanisms~\cite{Bonnet:2012kh}. In such case new signatures are
expected at colliders such as the
LHC~\cite{AguilarSaavedra:2012fu,Das:2012ii}.
The deep significance of \nbb rests upon the fact that, irrespective
of its origin, the observation of \nbb implies \lnv and the Majorana
nature of neutrinos~\cite{Schechter:1981bd,Duerr:2011zd}.
This summarizes our brief summary of the experimental status of
neutrino physics.  For the corresponding references, see
Refs.~\cite{McDonald:2016ixn,Kajita:2016cak},~\cite{Valle:2015pba,KamLAND-Zen:2016pfg}
and references therein. Now we turn to theory, starting with the
origin of neutrino mass and then moving to the structure of neutrino
mixing and their theoretical origin.\\[-.1cm]

\noindent
\textbf{\encircle{F}  Weinberg dimension five operator. }
The origin of neutrino mass is one of the most well kept secrets of
nature. A very general model independent approach was suggested by
Weinberg, who noted that one can form a dimension five
operator\footnote{It is also possible to induce neutrino mass through
  higher order operators~\cite{bonnet:2009ej}.} , with the \sm lepton
and Higgs doublets. This turns into a Majorana neutrino mass once the
electroweak symmetry breaks through the nonzero vacuum expectation
value (vev) of the Higgs doublet.
The smallness of neutrino mass would be ascribed to the mass scale
characterizing the \textit{d=5} operator~\O, which was originally
expected to be violated at high scale.
\begin{figure}[h]
\centering
\includegraphics[width=0.3\textwidth,height=2.5cm]{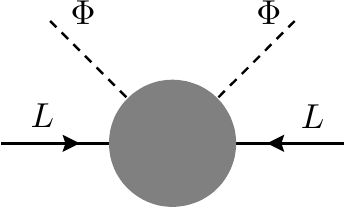}
\caption{Dimension 5 operator leading to neutrino mass.}
\label{d5}
\end{figure}
However, the operator \O may also be characterized by a small scale,
so that in its absence the symmetry of the theory would increase,
since lepton number would be recovered. This is a realization of
t'Hooft's naturalness criterion.
This is a generic argument. Nothing is known about the underlying
mechanism that engenders this operator, its characteristic scale or
its flavor structure. For that we need a theory. Various alternative
theories may be classified by the way they generate the operator \O,
with
two broad sub-categories, namely seesaw and radiative schemes.\\[-.1cm]

\noindent
\textbf{\encircle{G}  Standard seesaw mechanism. }
This is by far the most popular approach to neutrino mass
generation~\cite{Valle:2015pba}.  It assumes that the dimension five
operator arises at tree level either through the exchange of new heavy
right-handed neutrinos (type-I seesaw) or by the exchange of a triplet
of scalars (type-II or triplet seesaw).
Although the seesaw can be motivated by grand unified theories (GUTS)
or by models with intermediate scales (e.g. Pati-Salam or
Peccei-Quinn), the most general seesaw formulation is at the \sm
level~\cite{Schechter:1980gr}~\footnote{ It is curious that, in
  Ref.~\cite{Schechter:1980gr}, the type-I/type-II naming was swapped
  with respect to what became later established.}.
\begin{figure}[h]
\centering
\includegraphics[width=0.475\textwidth,height=2.5cm]{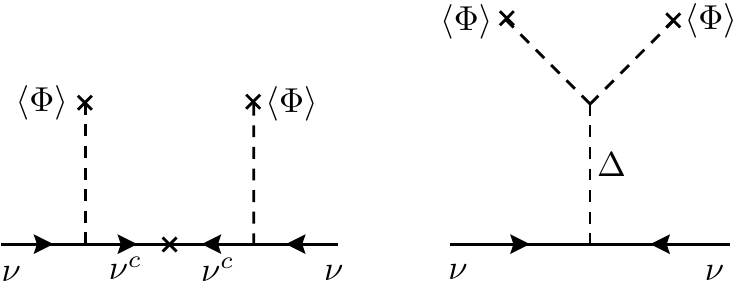}
\caption{Type-I (Left) and Type-II (Right) seesaw mechanism.}
\label{hseesaw}
\end{figure}
In its original formulation, for example, the type-I seesaw mechanism
was thought to involve the exchange of heavy intermediate fermions, at
a mass scale associated, say, with grand unification. Barring the the
use of arbitrarily small Yukawa couplings to account for the small
neutrino masses, there are no expected collider physics implications
in this case. See, in contrast, discussion on low-scale seesaw, below.\\[-.1cm]

\noindent
\textbf{\encircle{H}  Low-scale seesaw mechanism and t'Hooft's naturalness. }
One should realize that the seesaw mechanism is not a model, rather a
general framework for generating neutrino mass. As such, its general
formulation clearly allows for ``\textit{genuine}'' low-scale
realizations~\footnote{\textit{Genuine} low-scale means that tiny
  neutrino masses do not require arbitrarily small parameters, such as
  Yukawa couplings. The seesaw scale in any high-scale type-I seesaw
  can be made arbitrarily low by lowering the Dirac Yukawas.} so that
when the coefficient of the operator {\O} becomes zero the symmetry of
the theory enhances as a result. 
\begin{figure}[h]
\centering
\includegraphics[width=0.37\textwidth,height=2.4cm]{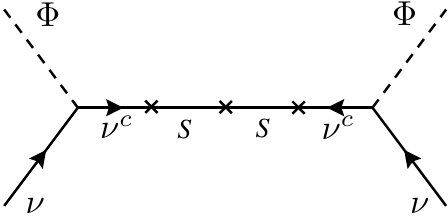}~~~~~~~~~~~~
\includegraphics[width=0.37\textwidth,height=2.5cm]{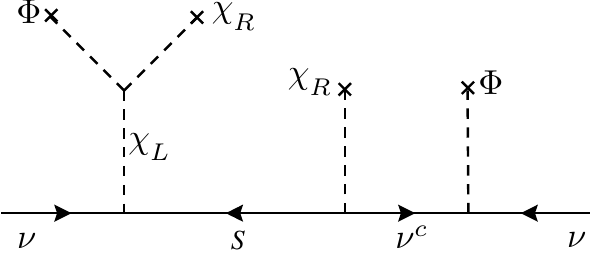}
\caption{Low-scale seesaw mechanism: inverse (left) and linear
  (right) realizations.}
\label{lseesaw}
\end{figure}
Hence there is, in this case, no need for the accompanying physics to
live at a large scale.  This is the theoretical basis of the low scale
seesaw mechanism.  There are two variants, namely linear and inverse
seesaw.  They are currently very popular as they open the possibility
of direct production of the neutrino mass generation messengers at
colliders experiments. In the presence of a new gauge portal these
lead to \lfv signatures at colliders such as the LHC or the future
proposals such as
ILC/CLIC~\cite{AguilarSaavedra:2012fu,Das:2012ii}.\\[-.1cm]

\noindent
\textbf{\encircle{L}  Dirac seesaw mechanism. }
Almost forty years after the seesaw idea first appeared, we have
developed a full conceptual description of both Type-I as well as
Type-II seesaw siblings of the seesaw mechanism for the case of Dirac
neutrinos.
In order to ensure the Dirac nature of neutrinos, two states
associated to ``left'' and ``right'' are needed and, moreover, some
extra symmetry principle is needed for ``Diracness''.
\begin{figure}[!h]
\centering
\includegraphics[width=0.35\textwidth,height=2.8cm]{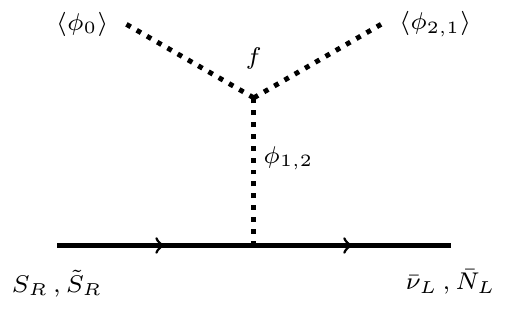}
\caption{Type-II Dirac seesaw mechanism.}
\label{d2seesaw}
\end{figure}
For example, in the type-II case the smallness of neutrino mass
follows from a parameter whose absence enhances the symmetry of the
theory, hence \textit{natural} in  t'Hooft's sense.
Neutrino mass generation may result from the spontaneous breaking of a
global U(1) symmetry, leading to a physical Nambu-Goldstone boson - a
Dirac sibling of the \textit{majoron} - we call \textit{Diracon}. Like
the majoron case also the \textit{Diracon} couplings are severely
restricted by bounds from stellar cooling rates in
astrophysics. Although stringent, these are consistent with possibly
significant invisible Higgs decays to \textit{Diracons}, well
constrained by studies at colliders such as the LHC.\\[-.1cm]

\noindent
\textbf{\encircle{M}  Radiative mechanisms. }
These constitute an interesting alternative to generate {\O}
without the need to invoke physics at inaccessible mass
scales~\cite{Boucenna:2014zba}.
\begin{figure}[!h]
\centering
\includegraphics[width=0.35\textwidth,height=3.5cm]{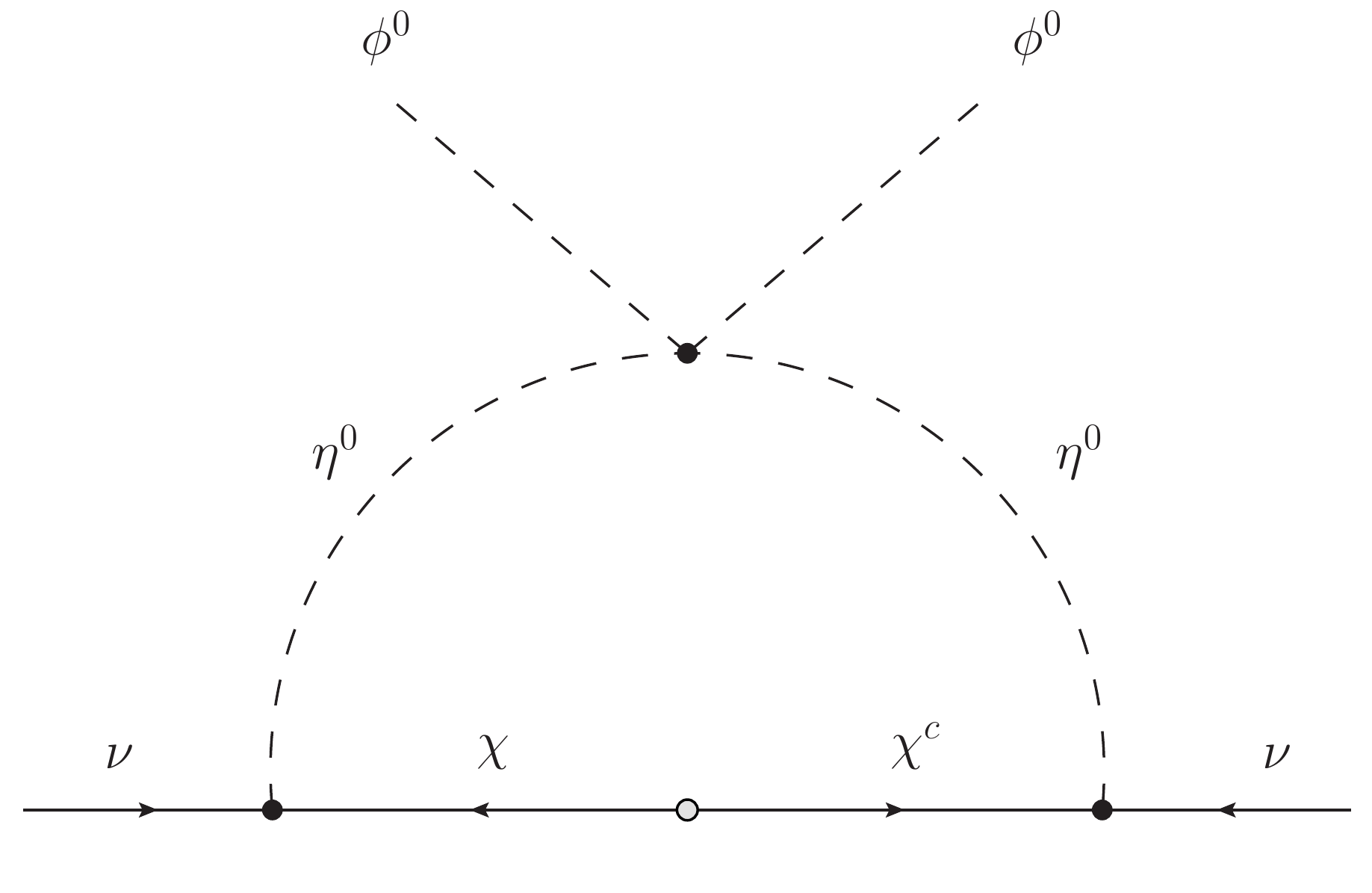}
\caption{Dark matter as messenger of radiative neutrino mass
  generation~\cite{Hirsch:2013ola,Merle:2016scw}.}
\label{scoto}
\end{figure}
These typically involve new scalar bosons. A specially attractive
example are the scotogenic models~\cite{Ma:2006km}, interesting
because they naturally incorporate dark matter. The latter emerges as
a messenger of neutrino mass generation. The model invokes a $Z_2$
symmetry ensuring the radiative nature of neutrino mass and, at the
same time, also stabilizing dark matter. A phenomenological richer
realization of Ma's original idea has been proposed and studied
in~\cite{Hirsch:2013ola,Merle:2016scw}.
It is also possible that, in extensions of the electroweak gauge
symmetry such as \TrTrOne schemes, neutrino masses may be induced
radiatively as a result of the exchange of new gauge
bosons~\cite{Boucenna:2014ela}.\\[-.1cm]

\noindent
\textbf{\encircle{N}  The flavor problem. }
The pattern of charged fermion masses is very strange, spanning about
six orders of magnitude between the electron mass and the top quark
mass.
This is just one aspect of the flavor problem.
Another one is the observed disparity between quark and lepton
mixings. In fact, as seen above, the smallest of the lepton mixing
angles, i.e. the ``reactor'' angle $\theta_{13}$, is similar to the
largest of the quark mixing angle, namely the Cabibbo angle.
In fact this may well be a subtle message nature is conveying to
us. For example there may be a symmetry rationale in which the
Cabibbo angle acts as a universal seed for all quark and lepton
mixings~\cite{Boucenna:2012xb,Ding:2012wh,Roy:2014nua}
Also the magnitudes of the lepton mixing angles do not seem arbitrary
parameters.
The way we face the challenge of bringing some rationale to the
observed pattern of fermion masses and mixings is through the use of
symmetry approaches. In this context let us mention the interesting
case of non-Abelian flavor symmetries. \\ [-.1cm]

\noindent
\textbf{\encircle{O}  Flavor-dependent b-$\tau$ unification. }
As an example we mention here a successful flavor generalization of the b-$\tau$
unification formula characteristic of minimal SU(5), namely,
\begin{equation}
  \label{eq:b-tau}
\frac{m_b}{{\sqrt{m_d m_s}}} = \frac{m_\tau}{{\sqrt{m_e m_\mu}}}
\end{equation}
This relation holds approximately in a number of flavor models based
on the $A_4$~\cite{Morisi:2011pt,King:2013hj,Morisi:2013eca} and
$T_7$~\cite{Bonilla:2014xla} symmetries. This formula is a consistent,
and fairly stable, generalization of the conventional SU(5)
prediction, showing how one can relate quark and lepton masses without
the need for grand-unification.\\[-.1cm]

\noindent
\textbf{\encircle{P}  BMV model as prototype flavor model. } 
We now turn to the minimal flavor scheme proposed
in~\cite{babu:2002dz}, we call it BMV, for short. The model adopts A4,
the smallest non-Abelian symmetry group with three-dimensional irreps
where the three families of leptons can fit nicely.  Valid at some
high-energy scale, the flavor symmetry naturally leads to degenerate
neutrino masses and hence to a sizeable rate for neutrinoless double
beta decay.
Realistic neutrino mass splittings are then induced by
renormalization group evolution with threshold corrections.
Assuming CP invariance the atmospheric mixing is predicted to be
maximal, $\theta_{23}=\pi/4$, and the reactor mixing $\theta_{13}$ is predicted to
vanish~\footnote{In the presence of CP violation $\theta_{13}$ is
  arbitrary and CP must be maximally violated in neutrino
  oscillations~ggg}. Note that \lfv processes such as
$\tau \to \mu \gamma$ are expected to lie
within reach of upcoming experiments~\cite{hirsch:a4}.
However, given the reactor results, e.g. from Daya-Bay, the prediction
$\theta_{13}=0$ needs to be corrected.\\[-.2cm]

\begin{figure}[!h]
\centering
\includegraphics[width=0.47\textwidth,height=4.5cm]{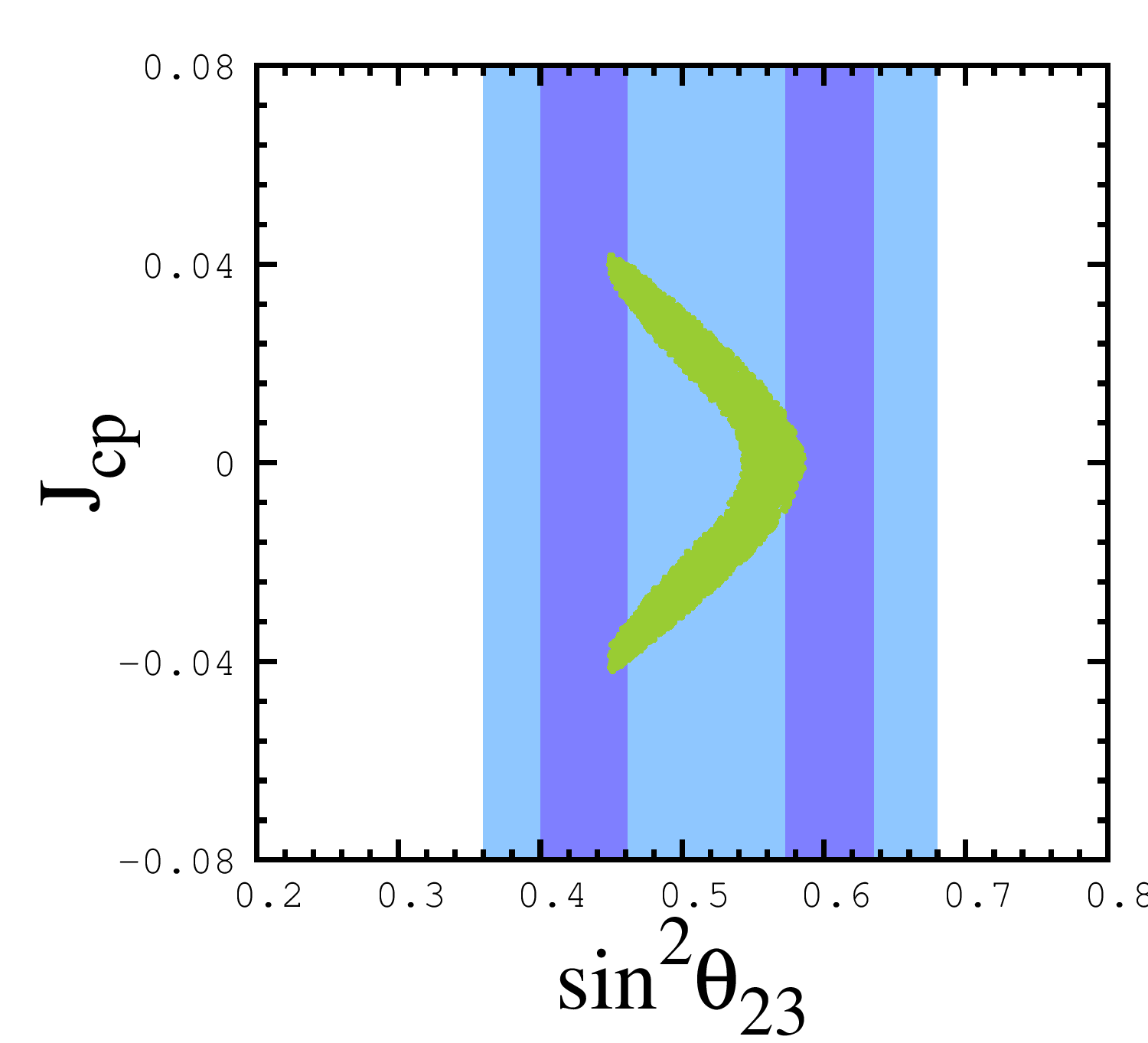}
\includegraphics[width=0.47\textwidth,height=4.5cm]{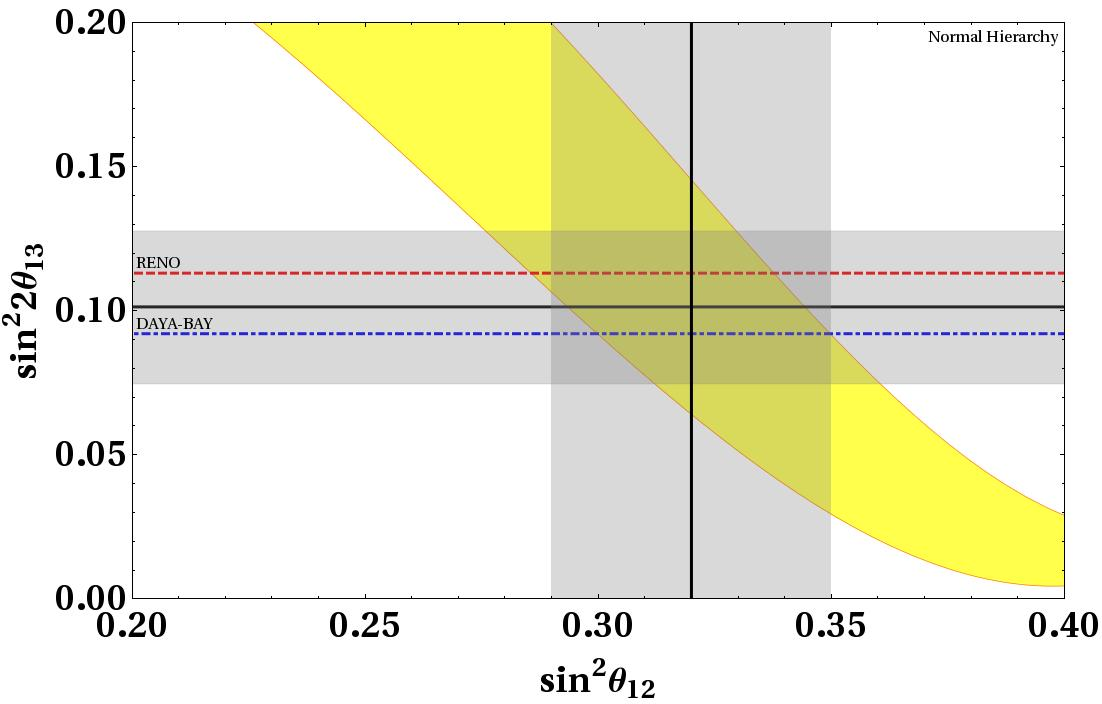}
\caption{Left: CP versus atmospheric angle correlation predicted in
  revamped BMV model~\protect\cite{Morisi:2013qna}. Right: Correlation
  between reactor and solar mixing in $\Delta(54)$ flavor
  model~\protect\cite{Boucenna:2012qb}. Bands indicate oscillation results.}
\label{ref:bmv}
\end{figure}

\noindent
\textbf{\encircle{Q}  Oscillation prediction in revamped BMV. }
Generalizing BMV is simple and leads to non-trivial results. One finds
that not only $\theta_{13}$ is generated, and hence CP violation in
oscillations, but also a departure from maximality in the atmospheric
mixing $\theta_{23}$. Moreover, that these parameters are correlated
as shown in Fig.~\ref{ref:bmv}, left panel.
The green region is the theory prediction, while the narrow dark
bands are 1$\sigma$ and the broad light band corresponds to
3$\sigma$, according to the global fit \cite{Forero:2014bxa}.
\newline \vskip -.7cm \noindent
In fact, very often the predictions one obtains from flavor models are
expressed as correlations between oscillation parameters, for example
between the solar angle $\theta_{12}$ and the reactor $\theta_{13}$,
as seen in the right panel of Fig.~\ref{ref:bmv}.
There is a huge literature on flavor model building, many symmetries
can be used, one-by-one on a trial-and-error
basis~\cite{Hirsch:2012ym,Morisi:2012fg,King:2014nza}.
It is fun, but can be rather tedious.\\[-.1cm]

\noindent
\textbf{\encircle{R}  Residual CP symmetries. }%
Before closing this discussion, let me comment on an alternative
model-independent approach is to exploit symmetries of the neutrino
mass matrix, such as generalized CP symmetries, irrespective of how
exactly they emerge within a particular model.
This way one can study flavor predictions in a model-independent
way. The prototype case is mu-tau parity symmetry and generalizations
thereof~\cite{Grimus:2003yn}. For example, a particular prediction in
the $\delta_{CP}$ versus $\theta_{23}$ plane has been studied
in~\cite{Chen:2015siy} and many more were analysed
in~\cite{Chen:2016ica}.
These may be eventually tested by the new generation of neutrino
oscillation experiments, such as NOvA, DUNE and T2HK.\\

%
\noindent
\textbf{\encircle{S}  Gauge coupling unification. }
Within the \sm gauge coupling unification is a ``near-miss'': when
extrapolated to high energies from their measured low-energy values,
the three gauge couplings almost meet together, but not quite. Gauge
coupling unification constitutes, indeed, an attractive argument
suggesting the existence of physics beyond the Standard Model.
\begin{figure}[!h]
\centering
\includegraphics[width=0.47\textwidth,height=4.2cm]{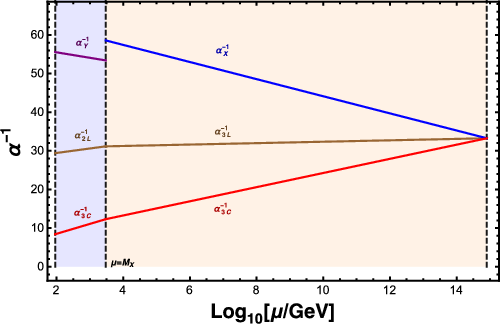}
\caption{Gauge coupling unification may be triggered by the physics
  inducing small neutrino mass~\cite{Boucenna:2014dia,Deppisch:2016jzl}. }
\label{ref:unif}
\end{figure}
\textit{What makes the gauge couplings unify?}  Common frameworks to
provide plausible answers to this question are (i) grand unification
and (ii) TeV scale supersymmetry. However, neither of their
characteristic predictions, such as proton decay or the existence of
supersymmetric states, have so far been vindicated experimentally.
A logical possibility is that gauge coupling unification is a
consequence of the same mechanism responsible for small neutrino mass
generation, as recently proposed~\cite{Boucenna:2014dia}.
This is illustrated in Fig.~\ref{ref:unif}.
The model employs an extended \TrTrOne electroweak gauge structure, in
the presence of a set of leptonic octets, directly involved in
neutrino mass generation~\cite{Deppisch:2016jzl}. Possible
embeddings probably require an F-theory GUT setup~\cite{King:2010mq}.\\[-.1cm]

\noindent
\textbf{\encircle{T}  What and where is the new physics? } 
So far there has been no hint of supersymmetry nor for any unexpected
signature. Given the lack of striking new results from the LHC a
burning question is what is the expected profile and scale of the new
physics? Can one expect any ``\textit{oasis}'' at the energies
currently available at the LHC?
Here we note that the historic discovery of the Higgs boson may
suggest that it is only the first of a family associated, perhaps, to
symmetry breaking patterns of more mundane extensions of the \SM gauge
group.
There are motivations for extended gauge structures. For example
left-right symmetry, which has as attractive feature the fact that it
elevates parity to the status of a spontaneously broken symmetry,
associated to the seesaw mechanism.
Likewise, there are chiral \sm extensions whose quantum consistency
requires exactly three families of fermions.
\textit{Which pattern of new physics should be expected at current
  and upcoming accelerator experiments?}  Similarly important,
\textit{On what
  grounds can one choose the preferred one?}\\[-.1cm]

\noindent
\textbf{\encircle{U}  Parity non-conservation, the number of families
  and new gauge bosons. }

\begin{figure}[!h]
 \centering
 \includegraphics[width=0.3\columnwidth,height=3.6cm]{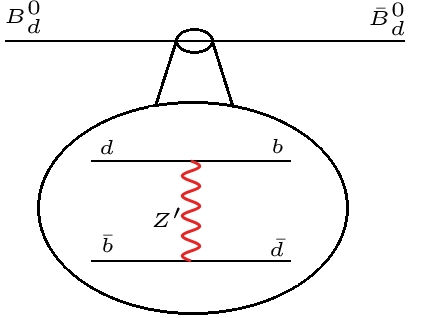}~~~~~~~
 \includegraphics[width=0.4\columnwidth,height=3.8cm]{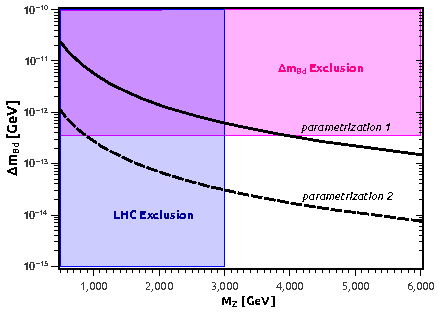}
 \caption{Feynman diagrams relevant for dilepton production at the LHC
   and the $B^0_d-\bar{B}^0_d$ mass difference in minimal low-scale
   \TrTrOne model.}
 \label{farinaldo}
\end{figure}

In order to provide a framework to answer the above questions
 one may consider some of the open conceptual issues in
weak interaction theory.
For example, (i) \textit{What is the role of parity
  non-conservation?}, (ii) \textit{what is the origin of neutrino
  mass?} or (iii) \textit{Is there a rationale for having just three
  families of quarks and leptons? } Here we consider these three
features as possible ways to establish a criterion for choosing the
\sm extension.
\begin{figure}[!h]
\centering
\includegraphics[width=0.45\textwidth,height=5cm]{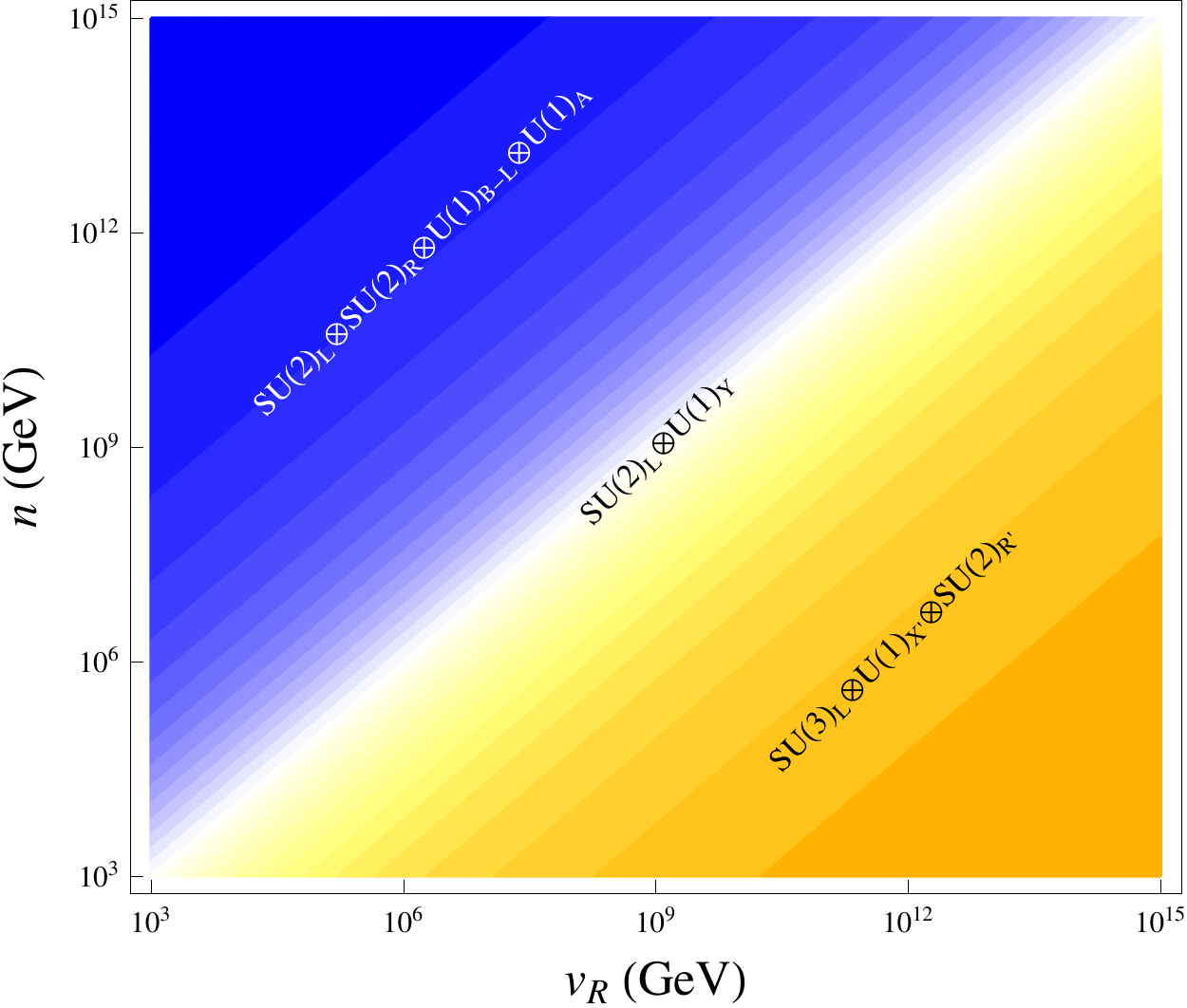}
\caption{Dynamically determining the new physics profile through the
  vev ratio  $n/{v_R}$~\cite{Reig:2016tuk}. }
\label{ref:3331}
\end{figure}
They can be reconciled in a left-right symmetric ``mother'' theory
based on the \TrTrTrOne gauge group~\cite{Reig:2016tuk}. Indeed, the
mother theory is a realistic manifestly left-right symmetric setup,
requiring the number of families to match the number of colors in
order to achieve quantum consistency.
Neutrino masses arise either from the canonical seesaw
mechanism~\cite{Reig:2016tuk} or from a low-scale seesaw
picture~\cite{Reig:2016vtf}.  Small neutrino mass correlates with the
observed V-A nature of the weak force.
Depending on the symmetry breaking path to the Standard Model one
recovers as the next step towards new physics either a conventional
\LR scenario or one based on a manifestly chiral extension of the
electroweak symmetry based on \TrTrOne in which the number of families
is fixed through anomaly cancellation.
If light enough, the resulting $Z^\prime$ gauge bosons can be produced
at the LHC, providing a production portal for the right-handed
neutrinos, whose decays would engender \lfv
processes~\cite{Deppisch:2013cya}.
On the other hand, its flavor changing interactions would also affect
the K, D and B neutral meson systems~\cite{Queiroz:2016gif}.
The interplay of the $B^0_d-\bar{B}^0_d$ mass difference (left panel
in Fig.~\ref{farinaldo}) and the LHC dilepton mass constraints for two
different parametrizations of the quark mixing
matrix is illustrated in right panel in Fig.~\ref{farinaldo}.
\begin{figure}[!h]
 \centering
 \includegraphics[width=0.3\columnwidth,height=3.2cm]{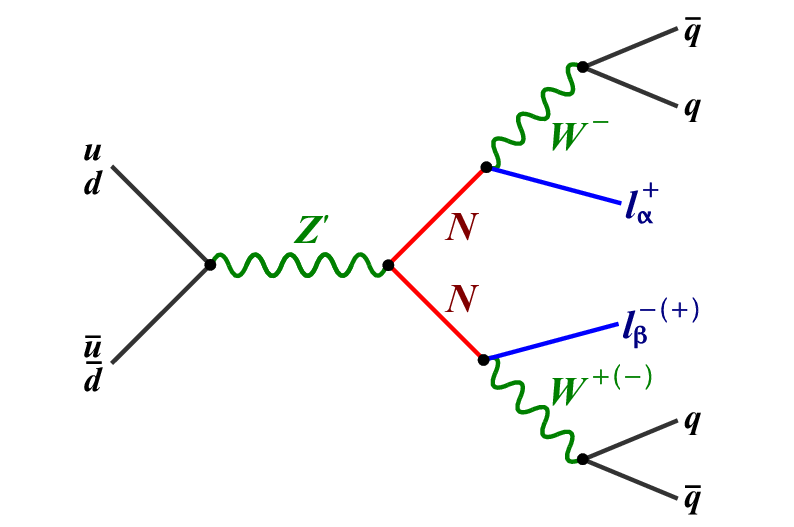}~~~~~
 \includegraphics[width=0.32\columnwidth,height=3.4cm]{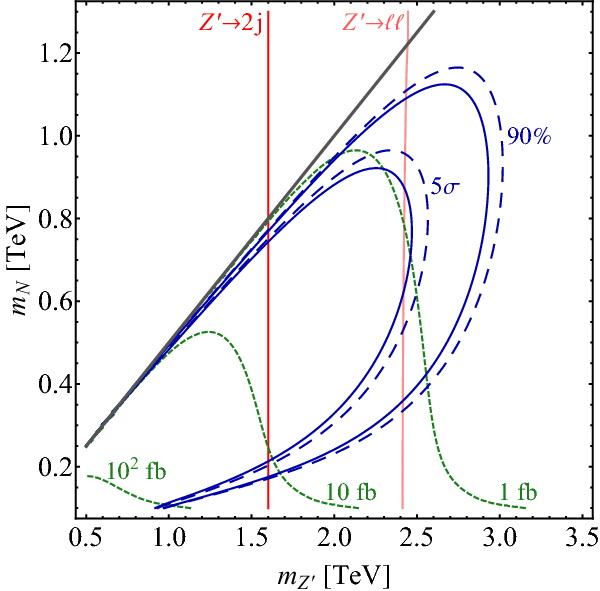}
 \caption{
   3-3-1 model.}
 \label{desai}
\end{figure}
In such scenario the right-handed neutrino messengers of neutrino mass
generation may be produced via a $Z^\prime$ portal and can decay via
small flavor violating couplings. This leads to \lfv at the
LHC~\cite{AguilarSaavedra:2012fu,Das:2012ii}, whose the rates are
unsuppressed despite unobservably small $\mu \to e+\gamma$ gamma
rates~\cite{Deppisch:2013cya}.\\[-.1cm]

\noindent
\textbf{\encircle{V}  Neutrino mass generation }
The historic discovery of the 125 GeV Higgs boson may suggest that it
is only the tip of the iceberg. Indeed, it is likely that this Higgs
is just the first of a family, there could be others associated with
the breaking of symmetries such as lepton number. For example extra
scalar multiplets beyond those in the \SM theory, such as singlet and
triplet Higgses, are used to generate small neutrino mass in the
seesaw mechanism. If ungauged, spontaneous \lnv implies the existence
of a physical Nambu-Goldstone boson - the
majoron~\cite{chikashige:1981ui,Schechter:1981cv}.
The good measurement of the invisible $Z^0$ decay width implies that
the majoron must be mainly singlet. 
If the associated scale is relatively low, such singlet ``majoron
model'' naturally implies potentially observable rates for invisible
Higgs decays~\cite{joshipura:1992hp}, leading to missing momentum
signatures at
accelerators~\cite{Bonilla:2015uwa,Bonilla:2015jdf}. Such are now well
constrained by LEP~\cite{Abdallah:2003ry} and LHC
experiments~\cite{Aad:2014iia}. \\[-.1cm]
\begin{figure}[!h]
\centering
\includegraphics[width=0.26\textwidth,height=2.55cm]{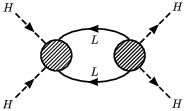}
\includegraphics[width=0.27\textwidth,height=2.6cm]{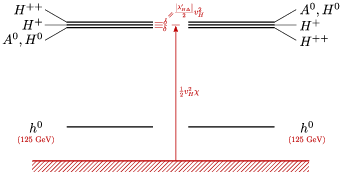}
\caption{ Small neutrino mass, vacuum stability and perturbative electroweak breaking. }
\label{ref:stability}
\end{figure}
\begin{figure}[!h]
\centering
\includegraphics[width=0.4\textwidth,height=4cm]{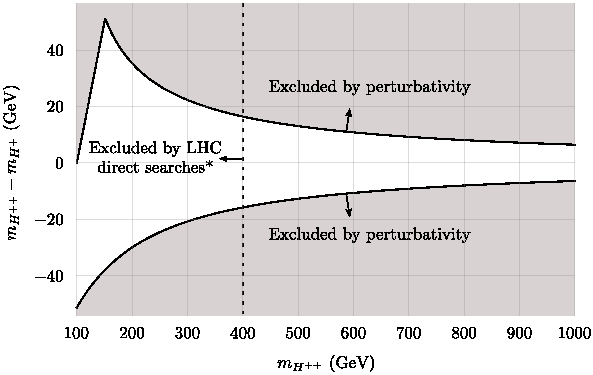}
\caption{Schematic type-II seesaw Higgs spectrum. }
\label{ref:stability}
\end{figure}
\\[-.1cm] \noindent
\noindent
\textbf{\encircle{W}  Neutrino mass generation and the consistency of the
  electroweak vacuum. }
We have just discussed that, if the smallness of neutrino mass is
associated to a spontaneously broken symmetry, we need extra Higgs
multiplets.
Hence, in addition to the \SM gauge invariance of the \sm we must also
break such symmetry, such as lepton number, so as to account for
naturally small neutrino mass.
In the presence of extra ``leptophilic'' scalars the quartic coupling,
whose positivity characterizes electroweak vacuum stability, gets new
contributions. This naturally provides a mechanism to stabilize the
theory's vacuum leading to a bounded-from-below potential energy.
~\cite{Bonilla:2015kna}. The mechanism may be \textit{qualitatively}
understood by ``squaring'' Weinberg' s operator, as indicated in
Fig.~\ref{ref:stability}.
If we now evolve the theory with renormalization group equations all
the way up to the Planck scale and require perturbativity to be
maintained, then we find further constraints on electroweak
breaking. For example, if a Higgs triplet is present, as in type-II
seesaw, the mass splitting of its components is strongly
restricted~\cite{Bonilla:2015eha}, leading to a ``compressed'' Higgs
mass spectrum, as seen in the middle and right panels in
Fig.~\ref{ref:stability}. They give a schematic view of the scalar
boson mass spectrum in the triplet seesaw model and one sees that the
heavy scalars are nearly degenerate. There is a broad class of
``neutrino motivated'' extensions of the \sm Higgs sector, which
provide interesting benchmark theories of electroweak
breaking.\\[-.1cm] \noindent

\noindent
\textbf{\encircle{X}  How about Gravity? }
As a fundamental interaction of nature, gravity is described
geometrically in Einstein's General Relativity. This is an elegant
classical theory, not part of the Standard Model. Indeed, we still
have not been able to develop a quantum theory of gravity.
Unifying gravity with the quantum field theoretic \sm description of
microphysics constitutes the biggest challenge in contemporary
physics.
The current approaches employ extra space-time dimensions either in
the context of string theories or of warped geometries.
Although both scenarios can be employed to provide frameworks for
adding gravity, here we just consider the effective four-dimensional
low energy theory that results from them after decoupling gravity. The
question one may pose is whether the latter can lead to some
predictions for neutrinos. For example, consider the extended \TrTrOne
electroweak symmetry framework discussed in \encircle{S}.  Although it
can not easily be unified within the conventional field theory sense,
it was shown that it admits a string completion within a quiver
setup~\cite{Addazi:2016xuh}. They constitute one of the most ubiquitous
particles in nature.  Thanks to their weak interaction, neutrinos are
excellent astrophysical messengers, probing the deep interior of the
Sun or of a supernova.  Likewise they probe the earliest instants of
the universe, just after the Big Bang.
Natural and artificial neutrino sources span about 20 orders of
magnitude in energy, all the way from the abundant neutrinos produced
in the Big-Bang to the ultra high energy cosmic ray neutrinos. The
former are abundant, though currently undetectable because of their
very low energy.  The latter have much higher interaction rates,
though their detection also constitutes a challenge thanks to the
small fluxes.
One finds that lepton and baryon numbers are perturbatively conserved,
so neutrinos are Dirac fermions.
Moreover, consistent anomaly cancellation require extra
``right-handed'' neutrino-like states which lead to a natural
realization of a novel, potentially low scale, type-II seesaw
mechanism~\cite{Reig:2016ewy}, as was illustrated in
Section~\encircle{L}, above.\\[-.1cm] \noindent

\begin{figure}[!h]
\centering
\includegraphics[width=0.55\textwidth,height=3.7cm]{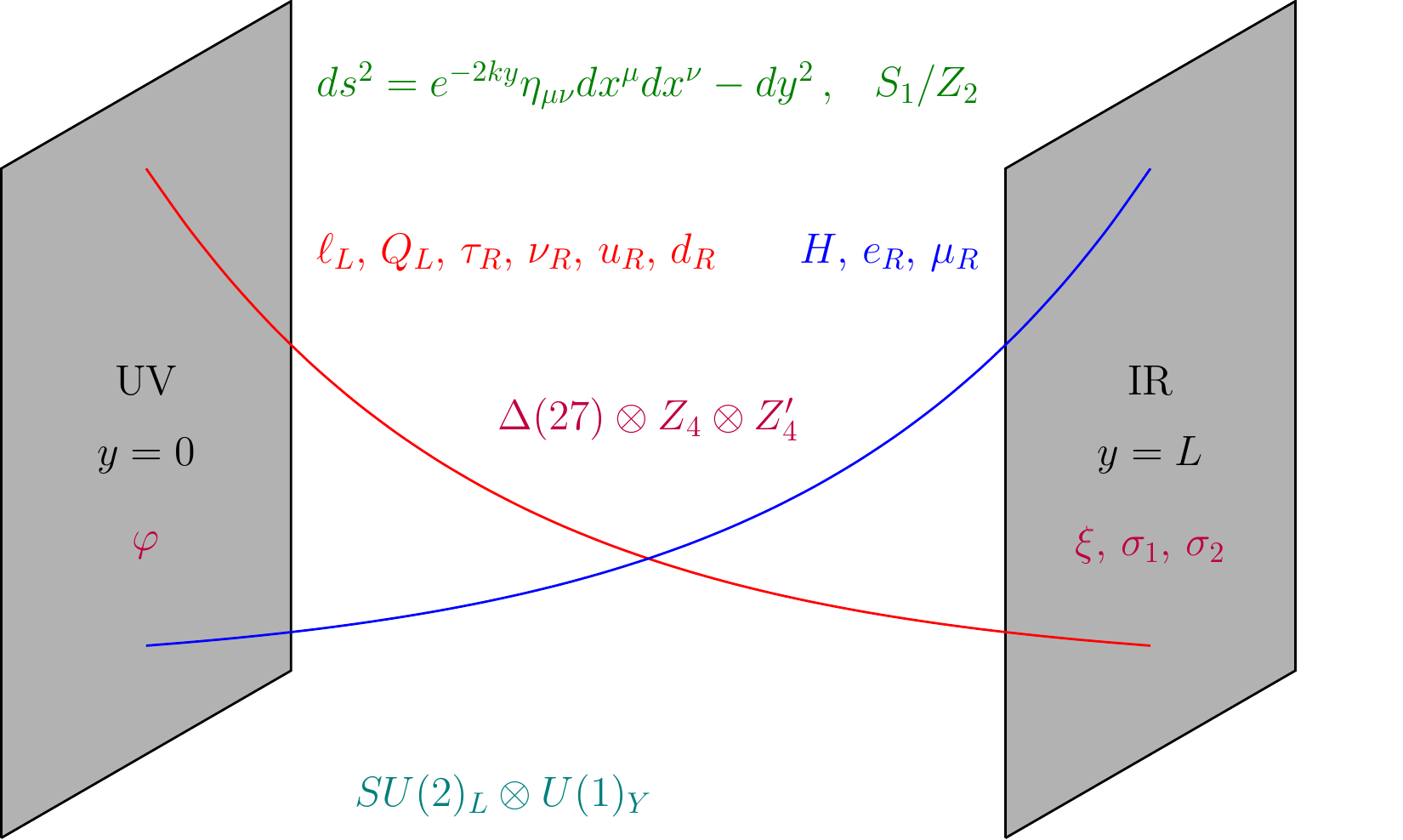}
\caption{Basic structure of the warped model, showing the UV (IR)
  peaked nature of the fields, from~\protect\cite{Chen:2015jta}. }
\label{ref:warped}
\end{figure}

\noindent
\textbf{\encircle{Y}  Neutrino predictions from Warped Standard Model. }
Randal and Sundrum~\cite{Randall:1999ee} suggested a
higher-dimensional mechanism to account for the hierarchy problem.
The weak scale is generated from a large scale of order the Planck
scale through an exponential hierarchy arising from the background
metric, assumed to be a slice of AdS5 space-time. 
In principle this nice mechanism may be used to ``explain'' other mass
hierarchies, such as those amongst the \sm fermions.
Ref.\cite{Chen:2015jta} proposed a realistic five-dimensional warped
\sm scenario with all fields propagating in the bulk, as illustrated
in Fig.~\ref{ref:warped}
\\[-.1cm] 
\noindent Mass hierarchies are accounted for by judicious choices of
the bulk mass parameters, while fermion mixing angles are restricted
by a $\Delta(27)$ flavor symmetry broken on the branes by flavon
fields.
Note that, like the other fermions, here neutrinos are Dirac type.
The flavor symmetry implies latter gives stringent predictions for the
neutrino mixing parameters, and the Dirac CP violation phase, all
described in terms of only two independent parameters at leading
order. This leads to a correlation between CP violation and the
atmospheric mixing parameter, in the same spirit as that in
Fig.~\ref{ref:bmv}. These lead to predictions for the upcoming long
baseline accelerator experiments T2K, NOvA and
DUNE~\cite{Pasquini:2016kwk}.
Note also that this Warped Standard Model
gives a realistic CKM matrix~\cite{Chen:2015jta}.\\[-.1cm]

\noindent
\textbf{\encircle{Z}  Neutrinos in Cosmology. }
According to the Big Bang the evolution of the universe started from a
very hot and dense past and went through a number of phase
transitions, dictated by the microphysics. In other words, the
particle physics describing the interaction of the elementary
constituents will determine how the universe evolves.
As one of the most ubiquitous particles in nature, neutrinos
play a key role in the evolution of the universe.
Thanks to their weak interaction, they can probe the earliest epochs
in the history of the universe, just after the Big Bang. In contrast,
through optical means the universe can only be probed after the
decoupling at 400,000 years or so~\cite{Ade:2015xua}.
Indeed, observations of temperature and polarization anisotropies of
the Cosmic Microwave Background (CMB) enable us to obtain limits of the
neutrino total mass.
However, neutrinos can probe much earlier epochs, such as that
associated with the breaking of the electroweak or higher symmetries.
Indeed, it is precisely in association new physics at these scales
that can address current puzzles in cosmology, such as dark matter,
inflation and generation of the baryon asymmetry.
One can formulate interesting new physics scenarios where the physics
associated with neutrino mass generation can potentially reconcile at
least some of the current cosmological puzzles as well. For example in
the scheme suggested in~\cite{Boucenna:2014uma} a single complex field
is added whose vacuum expectation value breaks lepton number and
generates neutrino mass through the seesaw mechanism, while the real
part drives inflation and the imaginary part plays the role of
metastable warm dark matter. Indeed this scenario has been shown to be
consistent with the CMB~\cite{berezinsky:1993fm,Lattanzi:2007ux} and
to lead to potentially viable detection by searching for X-ray
lines~\cite{Bazzocchi:2008fh,Lattanzi:2013uza}.\\[-.1cm]


\bigskip

\noindent
{\bf Acknowledgements: }
Supported by the Spanish grants FPA2014-58183-P, Multidark
CSD2009-00064, SEV-2014-0398 (MINECO) and PROMETEOII/2014/084
(Generalitat Valenciana).
I thank the school organizers for the warm hospitality and Carlos
Vaquera for reading the manuscript. 

\bibliographystyle{naturemag}

\end{document}